\documentclass[english]{article}
\usepackage[T1]{fontenc}
\usepackage[latin9]{inputenc}
\usepackage{geometry}
\usepackage{float}
\usepackage{amsmath}
\usepackage{graphicx}
\usepackage{natbib}
\makeatletter
\usepackage{xcolor}

\floatstyle{ruled}
\newfloat{algorithm}{tbp}{loa}
\providecommand{\algorithmname}{Algorithm}
\floatname{algorithm}{\protect\algorithmname}

\newcommand{\lyxaddress}[1]{
	\par {\raggedright #1
	\vspace{1.4em}
	\noindent\par}
}

\makeatother

\usepackage{babel}
\begin{document}

\title{{Drift Orbit Bifurcations and Cross-field Transport in the Outer Radiation Belt: Global MHD and Integrated Test-Particle Simulations}}

\author{{R. T. Desai$^1$\thanks{corresponding author: ravindra.desai@imperial.ac.uk}} , J. P. Eastwood$^1$, R. B. Horne$^2$, H. J. Allison$^3$, O. Allanson$^4$ \\ C. E. J.  Watt$^5$, J. W. B. Eggington$^1$, S. A. Glauert$^2$, N. P. Meredith$^2$, M. O. Archer$^1$ \\ F. A. Staples$^6$, L. Mejnertsen$^1$, J. K. Tong$^1$, J. P. Chittenden$^1$}

\date{}
\maketitle

\vspace{-2em}
\lyxaddress{\begin{center}
$^1$Blackett Laboratory, Imperial College London, London, UK
\par\end{center}}
\vspace{-3em}
\lyxaddress{\begin{center}
$^2$British Antarctic Survey, Cambridge, UK
\par\end{center}}
\vspace{-3em}
\lyxaddress{\begin{center}
$^3$GFZ German Research Centre for Geosciences, Potsdam, Germany
\par\end{center}}
\vspace{-3em}
\lyxaddress{\begin{center}
$^5$Department of Mathematics, University of Exeter, Penryn/Cornwall Campus; UK
\par\end{center}}
\vspace{-3em}
\lyxaddress{\begin{center}
$^5$Department of Mathematics, Physics and Electrical Engineering, Northumbria University, Newcastle Upon Tyne, UK
\par\end{center}}
\vspace{-3em}
\lyxaddress{\begin{center}
$^6$Mullard Space Science Laboratory, University College London, Surrey, UK
\par\end{center}}\vspace{-2em}

\begin{abstract}
Energetic particle fluxes in the outer magnetosphere present a significant challenge to modelling efforts as they can vary by orders of magnitude in response to solar wind driving conditions. In this article, we demonstrate the ability to propagate test particles through global MHD simulations to a high level of precision and use this to map the cross-field radial transport associated with relativistic electrons undergoing drift orbit bifurcations (DOBs). The simulations predict DOBs primarily occur within an Earth radius of the magnetopause loss cone and appears significantly different for southward and northward interplanetary magnetic field orientations. The changes to the second invariant are shown to manifest as a dropout in particle fluxes with pitch angles close to 90$^\circ$ and indicate DOBs are a cause of butterfly pitch angle distributions within the night-time sector. The convective electric field, not included in previous DOB studies, is found to have a significant effect on the resultant long term transport, and losses to the magnetopause and atmosphere are identified as a potential method for incorporating DOBs within Fokker-Planck transport models.
\end{abstract}

\section{Introduction}

The solar wind impinging on the day-side magnetosphere compresses the magnetospheric field lines and enhances the equatorial magnetic field magnitude \citep{Mead64,Northrup63,Roederer70}. Close to the magnetopause, the equatorial field strength can consequently exceed the field strength at a given particle's mirror point which causes drift orbits in the outer magnetosphere to bifurcate as particles become temporarily trapped within high-latitude pockets of magnetic field minima. Entering and exiting these non-dipolar regions has been associated with non-conservation of the particles second adiabatic invariant \citep{Antonova03} which, combined with conservation of the first adiabatic invariant, leads to radial transport across the magnetic field. It is therefore necessary to account for this phenomenon in order to understand and predict the dynamics of the outer radiation belt and its source populations.

The early works of \citet{Shabansky68} and \citet{Shabansky72} identified the phenomenon of drift orbit bifurcations (DOBs) and described how this can affect particles on both open and closed field lines. Early observations of enhanced energetic proton and electron fluxes in the high-latitude part
of the trapping region were identified by several early satellite missions \citep{Antonova73,Antonova79,Antonova96} which were consistent with later observations of cusp populations \citep{Savin98,Pissarenko01,Chen97,Chen98,Sheldon98}. These cusp populations were identified as being produced through tail injections \citep[e.g.][]{Baker98} which drifted around and become trapped in these high latitude pockets.

The changes to a particles second invariant were first outlined by \citet{Antonova03} and quantified through simulations by \citet{Ozturk07} who numerically integrated particle trajectories within symmetric super-imposed curl-free magnetic dipoles. These studies demonstrated how the instantaneous change to a particles pitch angle when undergoing DOBs is attributed to the particles orbit approaching and intersecting separatrix boundaries within Hamiltonian phase space \citep{Cary86} such that when the equatorial field intensity exceeds that at the mirror points, particles cross onto a new orbital contour \citep{Antonova03}. This conserves the total momentum but leads to a redistribution of the components parallel and perpendicular to the magnetic field.
The changes to the second invariant were found to be most significant for particles with small initial values, i.e. particles with pitch angles close to 90$^\circ$. A follow-up study by \citet{Wan10} looked at DOBs within asymmetric {\citet{Tsyganenko02b}} magnetic fields, imposed through a dipole tilt angle and a non-zero East-West component to the interplanetary magnetic field.  

The long-term evolution of electrons undergoing DOBs was examined by \citet{Ukhorskiy11} {within \citet{Tsyganenko07} fields} who showed how  particles can remain trapped and meander outward and inward within these non-dipolar regions for extended time periods. Through comparison with the magnetic diffusion rates of \citet{Brautigam00} they showed how DOBs can significantly enhance the recirculation and energisation of outer belt electron populations. \citet{Ukhorskiy14} subsequently examined DOBs within Tsyagenko fields subjected to upstream dynamic pressure variations on ultra-low frequency (ULF) timescales and showed they can lead to large increases in radial transport rates. The evolution of the distribution functions in the outer magnetosphere also notably deviated from the diffusion approximation. 

DOBs have been considered in relation to the calculation of the last closed drift shell \citep{Albert18} which presents an important outer boundary for  Fokker-Planck radiation belts models \citep[e.g.][]{Subbotin09,Glauert14b}. These typically use Tsyganenko field models for calculation of the third invariant as well {as the location of the last closed drift shell} but, as the third invariant is undefined for a bifurcating orbit, there exists a region of the outer magnetosphere where the particle dynamics are not fully unaccounted for. Global magnetohydrodynamic (MHD) simulation codes provide an alternate method of modelling the global magnetic field and test-particle simulations within global MHD fields have yielded valuable insights into magnetospheric particle dynamics \citep[e,g,][]{Hudson97,Elkington02,Kress07,Sorathia17}. The analysis of DOBs within global MHD fields also allows the inclusion of self-consistent magnetospheric electric fields not considered in previous studies of DOBs. 

In this study we use the global MHD model, Gorgon \citep{Mejnertsen2018,Eggington2020,Desai21}, and integrated test-particle simulations to demonstrate that this modelling approach is capable of capturing and further studying DOBs. Section \ref{GMHD} first describes the global MHD simulations and outlines the resolution requirements of $\approx$1/4 R$_E$ as necessary to properly resolve the non-dipolarity in the locally enhanced magnetic fields near the sub-solar point.  Section \ref{TP} then describes the test-particle simulations and how, through a combination of tracing particle trajectories and magnetic field lines, the simulations are able to conserve and measure a given particle's adiabatic invariants to a high level of precision. Ensembles of energetic electrons are subsequently simulated traversing the bifurcation region for southward and northward Interplanetary Magnetic Field (IMF) orientations and the differing transport in each scenario constrained as a transport map of equatorial pitch angle and radial location. The simulations agree in predicting large increases in the second invariant for particles with pitch angles close to 90$^{\circ}$, which is suggested as a mechanism which produces butterfly pitch angle distributions within the night sector. The long-term evolution is then examined and the simulations are found to depend heavily upon the convective electric fields, and DOB-driven losses to the magnetopause and atmospheric loss cone are suggested as potential effect{s} which can be incorporated within diffusive radiation belt models. 
  
\section{Global-MHD Simulations}
\label{GMHD}
\subsection{Method}

The resistive MHD simulation code, Gorgon, was developed for the study of high-energy-density laboratory plasmas \citep{Chittenden2004,Ciardo2007} and has since been adapted to the study of the solar wind-magnetosphere interaction \citep{Mejnertsen2016,Mejnertsen2018,Eggington2020,Desai21}. Gorgon is differentiated from other global MHD codes in that it solves for the vector potential on a staggered grid and automatically conserves the divergence of the magnetic field, $\nabla \cdot B = 0$, to machine precision without the need for an additional divergence cleaning operation. Gorgon employs a Message Passing Interface (MPI) domain decomposition parallelisation and has evolved to include higher order solvers and a thin shell ionosphere which maps region 1 field aligned currents to and from the ionosphere \citep{Eggington2018}. 
The MHD equations are solved on a cartesian grid extending from -20 to 130 R$_E$ in X with the Sun to the left at --X, and -40 to 40 in Y and Z.  In this study we examine the magnetosphere under the influence of both southward and northward steady solar wind conditions of: v$_x$ = 400 km/s; T$_i$=T$_e$=5 eV, proton density n$_p$=5 cm$^{-3}$ and with B$_z$ = -2 and 2 nT. This yields a dynamic pressure of 1.34 nPa and produces a sub-solar magnetopause stand-off distance of --10.75 and --11.25 R$_E$ for the two respective IMF orientations.

\begin{figure*}[ht]
\hspace{-2em}
\includegraphics[width=1.1\textwidth]{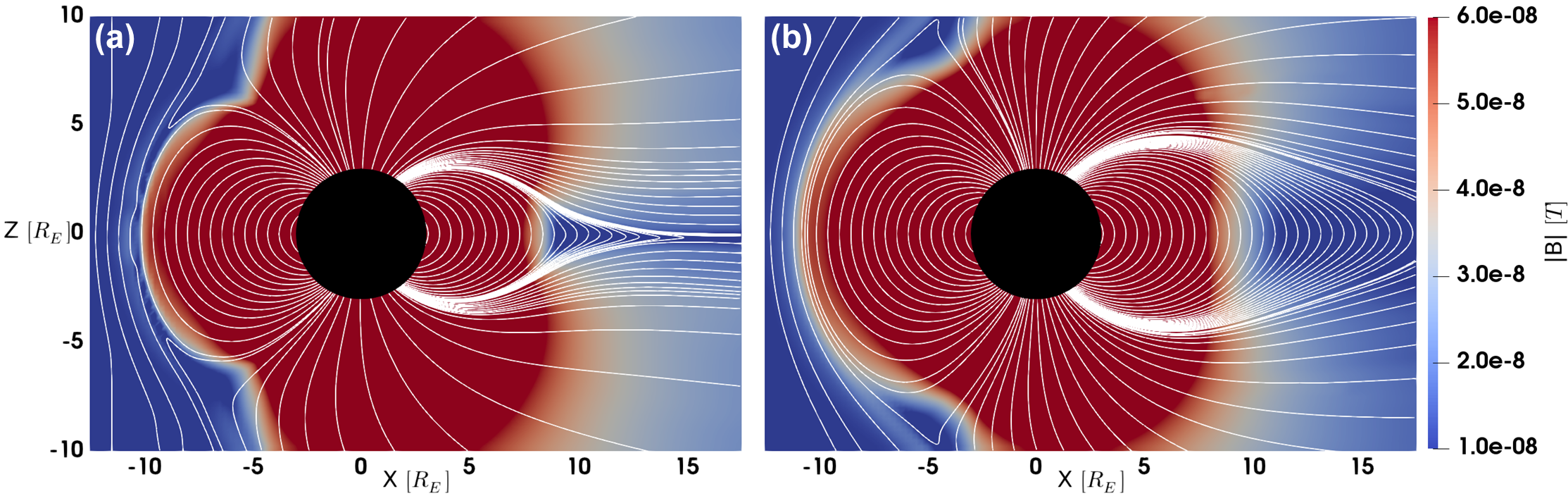}
\caption{Global magnetic field configurations in the X-Z plane for a grid resolution of 1/4 R$_E$. Panels (a) and (b) correspond to southward and northward-oriented IMF conditions, respectively.  The colour scale is optimised to capture high-latitude minima along the outer day-side field lines.
\label{MHD}}
\end{figure*}  

 \subsection{Magnetic Minima}
 
 Figure \ref{MHD} shows a slice through the two simulations run with a grid resolution of 1/4 R$_E$ and zero dipole tilt, two hours after initialisation. The two panels correspond to southward and northward IMF conditions. The oppositely directed IMF produce several differences to the configurations of the global fields. The most obvious of these is that southward IMF induces reconnection at the sub-solar point as evidenced by field lines being dragged poleward whereas the northward IMF instead induces reconnection at high latitudes. This effect slightly erodes the magnetopause such that the sub-solar magnetopause boundary is located $\approx$0.5 $R_E$ further inward {for IMF B$_z<$0} and results in different global convection patterns and a visible near-Earth neutral line in the magnetotail just beyond X $\approx$ 15 R$_E$. %
 The colour scale is limited to between 10 and 65 nT to show the variations in the magnetic field strength along field lines close to the magnetopause. Selected fields lines are traced in white which reveals the anticipated equatorially enhanced magnetic field and the minima in the B field at high latitudes. For example, at X = 8 R$_E$ in Figure \ref{MHD}a, the magnetic field is $\approx$20 nT greater at the equator (Z=0) then the cusps (Z = $\pm$ 4 R$_E$).
 
 Magnetospheric dynamics occur on a variety of scales and even large-scale magnetospheric phenomena, such as ultra-low-frequency pulsations, have been shown to be highly dependent on the selected grid resolution \citep[e.g.][]{Claudepierre09}. There consequently exists a trade-off between resolving the phenomenon of interest and a tractable computational problem. \citet{Kress07} show DOBs occurring within combined MHD and test particle simulations using the LFM global MHD model with a non-uniform grid but do not study this phenomenon further and we therefore first examine increasingly refined grid resolutions to determine the resolution necessary for resolving DOBs.
 
\begin{figure*}[ht]
\centering
\includegraphics[width=1.0\textwidth]{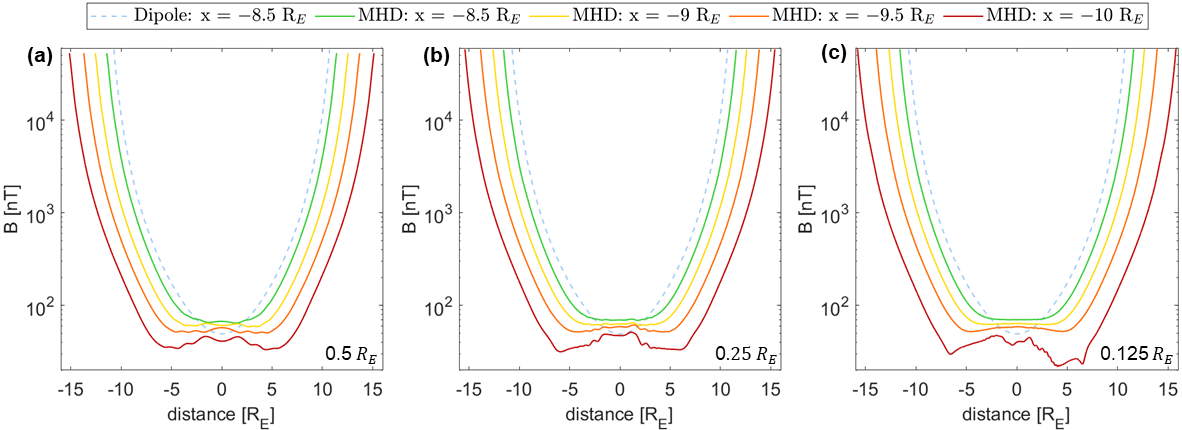}
\caption{Magnetic field strength along selected field lines which map to X = 8.5, 9, 9.5 and 10 R$_E$ upstream of the Earth. A dipole field mapping to 8 R$_E$ upstream is shown for comparison.
\label{convergence}}
\end{figure*}

Figure \ref{convergence} shows the magnetic field strength along selected field lines at 0.5 R$_E$ intervals near the magnetopause for three separate simulations with grid resolutions of 1/2, 1/4 and 1/8 R$_E$ respectively, each run with a southward IMF. In all cases near the magnetopause the magnetic field displays a local maximum at the equator and two near-symmetric minima on either side of this at higher latitudes. This local maximum is comparable to those produced within the semi-empirical Tsyganenko field models \citep[see ][Fig. 1 therein]{Wan10}. This bifurcation region extends up to $\approx$ 1.5 R$_E$ inward from the magnetopause boundary and appears to diminish in intensity with decreasing radial distance.
This feature appears qualitatively similar at each resolution but is less pronounced in the lowest resolution case where the simulation under-estimates the extent of the minima along the outermost field line. This leaves a region of $\approx$ near the magnetopause where DOBs will not be accurately capture and due to this, and further test-particle simulations (not shown), we judge that the 1/4 R$_E$ resolution has converged and this resolution is selected moving forward. 

Oscillations in the magnetic field along the outer field-lines are also evident. These are caused by the southward interplanetary magnetic field (IMF) inducing reconnection at the subsolar point which means that flux rope structures are forming in this region. In the 1/8 R$_E$ resolution simulations, the outer-most field lines are also visibly in the process of reconnecting and being dragged poleward by the convecting solar wind, but this does not mean the resolutions disagree about the average minima.

\section{Test-Particle Simulations \label{sec:Parallel} }
\label{TP}
 \subsection{Method}
 
 In this study we trace electron trajectories within the self-consistent MHD fields using the test-particle approximation. The particles thus see the magnetic and electric fields but the particles' currents do not feedback into the global fields. The MHD simulations utilise a split dipole implementation \citep{Tanaka1994} which allows direct interpolation of the non-dipolar component to which the analytical curl-free dipolar component is added. We use the quadratic triangular shape cloud scheme \citep{Hockney81} to map between the fields and the particles' location sub-grid and clean spurious parallel electric fields \citep{Lehe09} which are zero within the ideal MHD approximation.  The test-particle simulations are implemented directly into the Gorgon code via a linked list data structure \citep{Tong2019} 
 and a vectorised MPI algorithm balances this additional load across all available processors. Previous studies have highlighted that in the magnetosphere, electron trajectories can generally be approximated by the guiding centre equations of motion \citep[e.g.][]{Elkington04} but particles close to the magnetopause can undergo current sheet scattering in which case finite gyroradius effects will become significant. We therefore integrate the full Newton-Lorentz equations of motion using the relativistic Boris integration scheme \citep{Boris70,Desai21a}. The particles are allowed to orbit anywhere within a radial sphere of 12 R$_E$ and can enter the MHD inner boundary at 3 R$_E$ at which point the MHD equations are no longer solved and the particles see a dipole field up until an atmospheric loss condition 1,000 km above the Earth's surface.

 To isolate the effects of drift-orbit-bifurcations from other magnetospheric phenomena we push the test-particles through individual MHD timesteps. The fields are therefore static but contain self-consistent electric fields resulting from plasma convection. We do this for multiple MHD timesteps obtained during constant solar wind conditions {following 2 to 2.5 hours of initialisation with a geomagnetic dipole moment of M = 7.94$\cdot$10$^{22}$ A m$^2$}. This therefore removes natural variations associated with magnetopause reconnection \citep{Russell96} and oscillations of the magnetopause surface \citep{Guo10} and any subsequent large-scale oscillations too.

\subsection{Adiabatic invariants}
 
 Constraining particle transport within the simulations requires high precision determination of a particles' adiabatic invariants.  The first adiabatic invariant is expressed as,
 \begin{equation}
     \mu = \frac{\gamma m_0 v_{\perp}}{2B},
 \end{equation}
 where $\gamma$ is the relativistic Lorentz factor, m$_0$ the rest mass, and v$_{\perp}$ the velocity perpendicular to magnetic field, B. This is trivial to calculate and is available at every location along a particle's orbit, either within the simulation or via post-processing. The second adiabatic invariant,
   \begin{equation}
     J = m_0 \oint \gamma v_{\parallel} \cdot dl,
 \end{equation}
 where $v_{\parallel}$ is the velocity parallel to B, is, however, more complicated to calculate.
 The discretisation of each particles trajectory into thousands to millions of steps mean it is not efficient to output every single datapoint and we therefore evaluate this integral as the particles trajectory is evolved and output diagnostic data including J/2, at a given particles mirror point. The third adiabatic invariant is defined as,
    \begin{equation}
       \Phi = \oint_c A \cdot dl = -\frac{2\pi B_0 R^2_E}{L^*},
       \label{phi}
 \end{equation}
where A is the vector potential. As shown, this is also commonly expressed in invariant coordinates where B$_0$ is the magnetic field strength at the magnetic equator on the Earth's surface, and L$^*$ is the particles L shell were all non-dipolar components slowly turned off. The third adiabatic invariant expressed in Equation \ref{phi} is, however, undefined for a bifurcating drift shell \citep[e.g.][]{Ozturk07,Ukhorskiy11} which is problematic for calculating radial transport rates. We therefore calculate radial transport rates in terms of dipole L shell under the assumption $\Delta$ L $\approx$ $\Delta$ L$^*$ at a given location in the magnetosphere and leave the problem of converting between physical and invariant coordinate systems for a further dedicated endeavour. 

Despite this assumption, measuring radial changes is not straight-forward as L has to be measured at a common location, before and after a drift orbit and a particle will in all probabilities arrive at the same local time with a different bounce and gyrophase. To account for this we offset the particle's gyroradius, and trace the gyro-centre location along the magnetic field down to the magnetic equator for both the initial and final positions.

  \begin{figure*}[ht]
\hspace{-2em}
\includegraphics[width=1.0\textwidth]{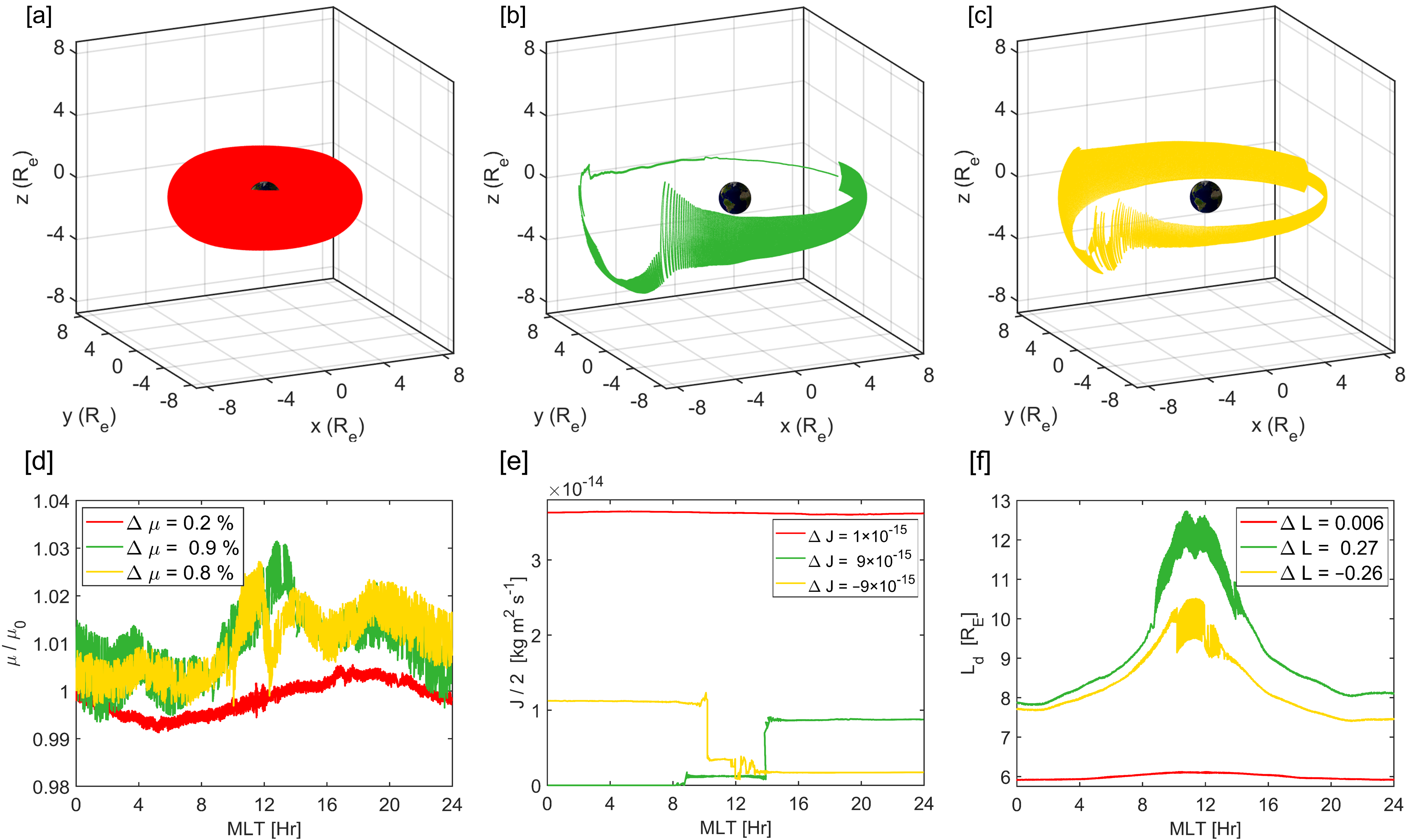}
\caption{Three MeV electron trajectories undergoing (a) a stable drift orbit at L=6,  (b) a bifurcated orbit which leads to an increase to the particles pitch angle and (c) a bifurcated orbit which leads to a decrease to the particles pitch angle. Lower panels (d--f) show the corresponding first and second adiabatic invariants and dipole L shell where the legends denote the change across the entire drift orbit.
\label{orbits}}
\end{figure*}  
 
 \subsection{Drift Orbit Bifurcations} 
 \label{DOB}
 
Figure \ref{orbits}a--c shows a selection of three 1 MeV electron orbits which represent a stable equatorially mirroring particle and two particles undergoing DOBs. Figure \ref{orbits}a shows the full orbit of the first electron which has an equatorial start position of L = 6 R$_E$ downtail and initial equatorial pitch angle, $\alpha_{0}$ = 45$^\circ$. The electron orbits anti-clockwise in MLT and mirrors to slightly higher latitudes on the day-side due to the compression of the magnetosphere before returning to its initial location. Figure \ref{orbits}b shows the second electron initialised at L = 8 R$_E$ downtail with $\alpha_{0}$ = 90 $^\circ$. This electron drifts along the magnetic equator into the bifurcation region on the day side and is deflected away from the local maximum at around 10.30 MLT and becomes trapped beneath the equator. The pitch angle increases significantly as it starts mirroring within the minimum B pocket in the southern hemisphere and, upon exiting this region near 13:30 MLT, the pitch angle increases further and the electron once again mirrors about the magnetic equator as it drifts back to the tail. Figure \ref{orbits}c shows the third electron starting at approximately the same radial distance as the second electron but this time with $\alpha_{0}$ = 65 $^\circ$. This third electron similarly avoids the local equatorial maximum on the day-side but this time its pitch angle decreases as the particle becomes trapped within the minimum B pocket and decreases even further upon exiting. 

Figure \ref{orbits}d-e displays the corresponding first and second invariants for these particle trajectories together with the particles L shell to represent radial transport. In all cases the first invariant remains approximately constant with a overall change from start to end of less than 1 \%. In the case of the stably trapped particle shown in Figure \ref{orbits}a, a sinusoidal variation is apparent due to the particle's interaction with the dawn-dusk electric field. This effect is also visible in the two bifurcating particle orbits shown in Figure \ref{orbits}b-c, although this field displays some modulation at noon close to the magnetopause boundary.

The second invariant is well-conserved for the stably trapped electron but undergoes several  changes for the bifurcating orbits. The changes to the pitch angles noted in Figures \ref{orbits}b--c, correspond to large step changes in J. This is also not constant within the trapped region and undergoes further variations which are attributed to the use of dynamic MHD fields and realistic local variations and inhomogeneities. 
The second invariant changes, or lack thereof,  are mirrored in the electrons' radial position due to the local conservation of the first invariant. The stably trapped electron consequently returns to a radial position less than 0.006 R$_E$ from its initial position. This small discrepancy could be attributed to the presence of a non-zero electric field but also represents the limit of our accuracy in measuring radial transport. The bifurcating drift orbits, however, do not return to the same radial position. Figure \ref{orbits}f shows how the two instantaneous increases to the second invariant corresponds to net outward radial transport of approximately 1/4 R$_E$ and how a decrease in the second invariant corresponds to net inward radial transport of the same amount. The L shell parameters vary significantly along the drift orbit due to the compression of the global field and further highlights why we measure radial transport at a common MLT.

It should be noted that in the DOB trajectories shown, the electrons either exhibit two increases or two decreases in J. This is not always the case and some particles exhibit an increase followed by a decrease in J, or vice versa, upon entering and exiting the minimum B pockets. This predominantly occurs for intermediary pitch angles, as discussed in the preceding Section \ref{mapping}, and is dependent upon the bounce phase which controls the location at which the particle enters the minimum B pocket.

\subsection{Transport Maps} 
\label{mapping}

To further constrain this radial transport mechanism, we derive maps for an ensemble of 1 MeV electrons initialised in the tail. The distribution covers a radial range of L = 6--10 in steps of 0.05, pitch angles, $\alpha_0$ = 30--150$^\circ$ in steps of 1$^\circ$, and an MLT range of 0:00 -- 00:02 minutes in steps of 10 seconds to resolve the electron bounce-drift period. The particles are traced through the magnetosphere for both southward and northward IMF conditions where the dynamic pressure is held constant, in order to examine effects associated with the differing structures of the reconnecting magnetopause boundary. The electrons are traced through a single drift orbit, as displayed in Figure \ref{orbits}.

\begin{figure*}[ht]
\centering
\includegraphics[width=0.7\textwidth]{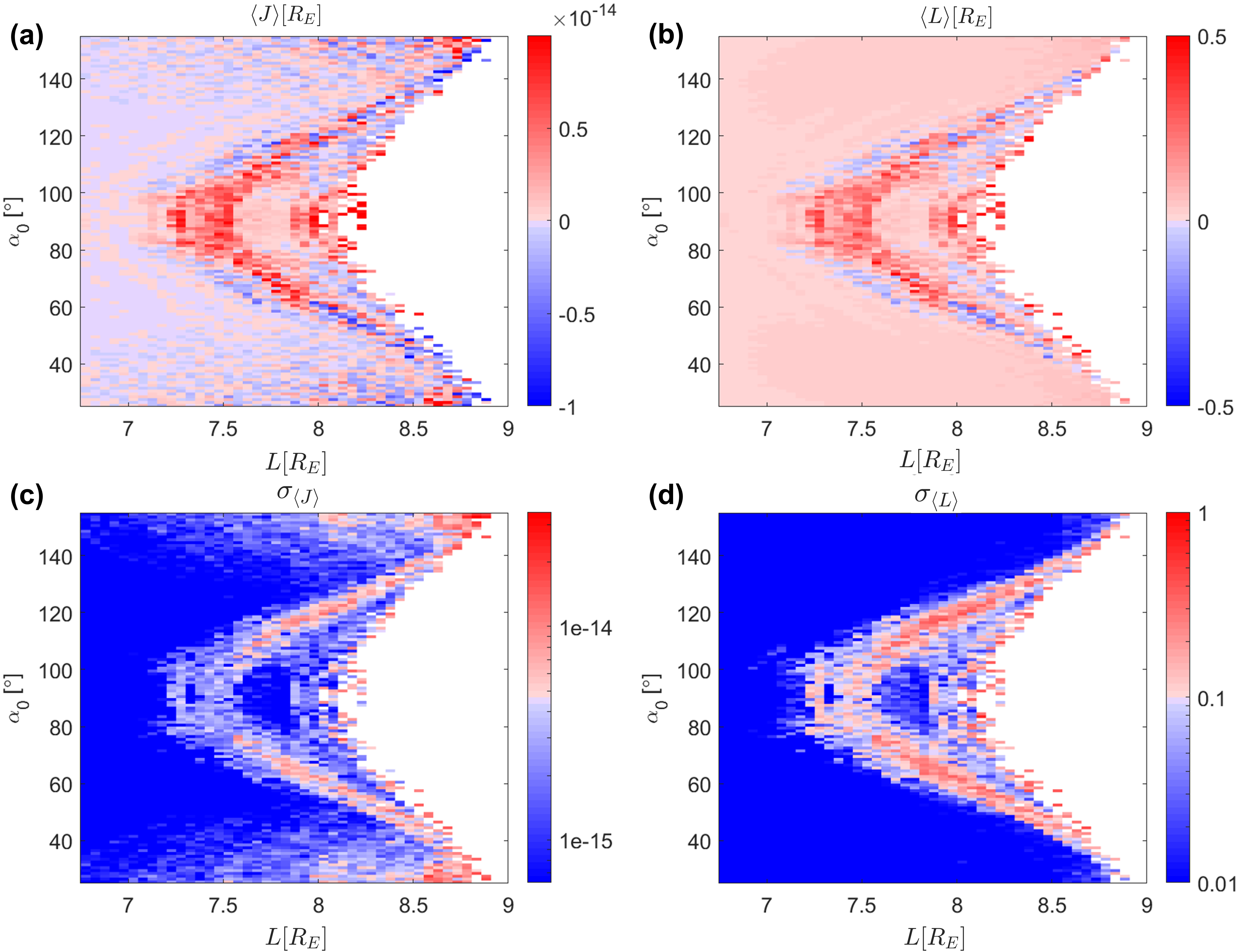}
\caption{Transport maps for southward IMF conditions. (a) shows $\Delta \bar{J}$ and (b) shows $\Delta \bar{L}$, both as a function of equatorial pitch angle $\alpha_0$ and initial position r$_0$ which is equivalent to dipole L shell. Lower panels (c) and (d) show the standard deviation, $\sigma$, over the multiple bounce phases.
\label{Smap}}
\end{figure*} 
 
\begin{figure*}[ht]
\centering
\includegraphics[width=0.7\textwidth]{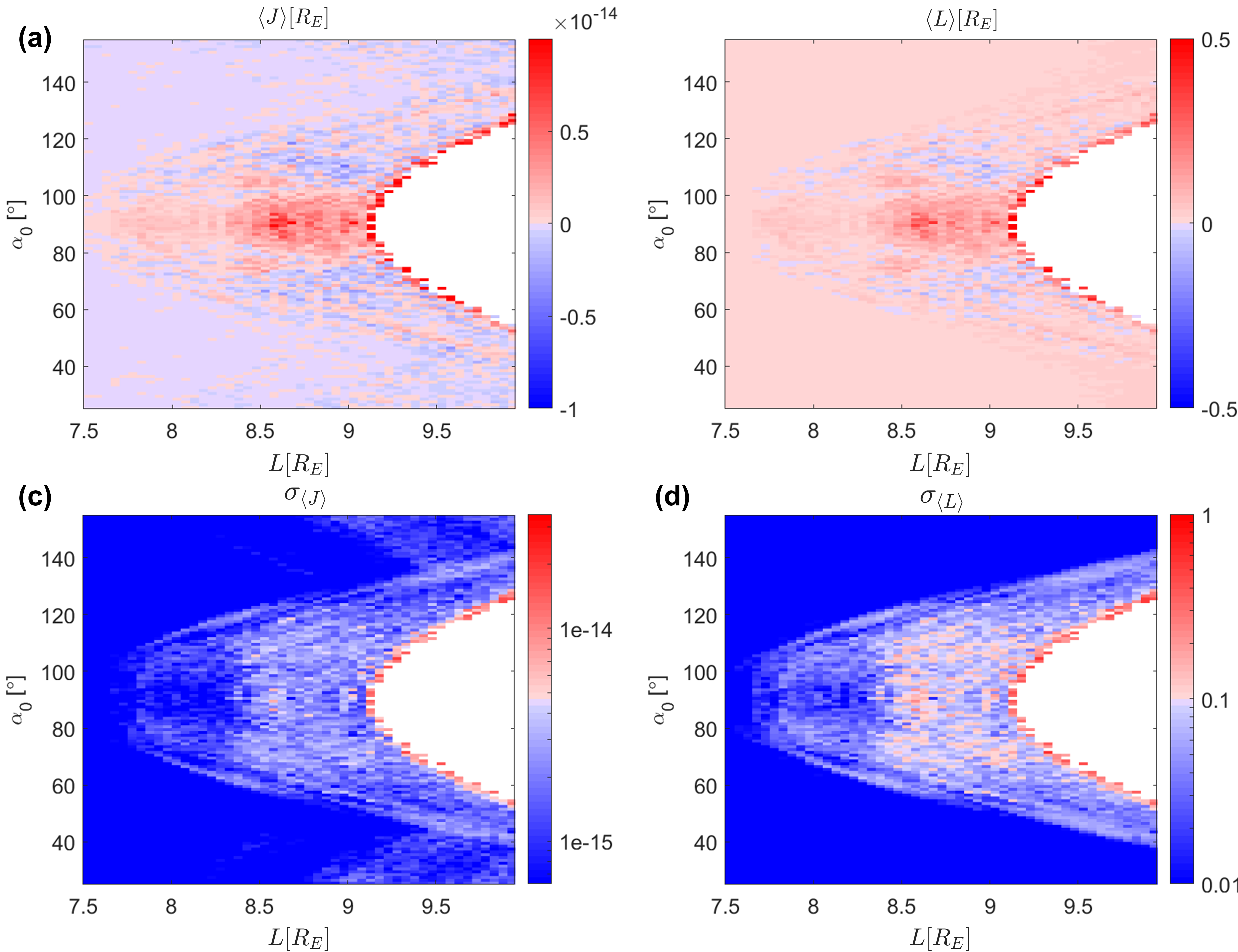}
\caption{Transport maps for northward IMF conditions. (a) shows $\langle \Delta J \rangle$ and (b) shows $\langle \Delta L \rangle$, both as a function of equatorial pitch angle $\alpha_0$ and L shell. Lower panels (c) and (d) show the standard deviation, $\sigma$, over the bounce phase distribution.
\label{Nmap}}
\end{figure*}

 Figures \ref{Smap} and \ref{Nmap} show the results of these two simulations for southward and northward IMF orientations, respectively.  The mean change of the ensemble, $\langle J \rangle$ and $\langle L \rangle$, are shown together with the standard deviation, $\sigma$, of electrons with different bounce phases. At larger L shells, the resultant transport maps show a drop-out in electrons which encounter the magnetopause loss cone which is higher for particles mirroring further from the equator. For large regions of the maps at lower L shells, $\langle J \rangle$ and $\langle L \rangle$ are approximately zero which represents negligible transport and stably trapped electrons analogous to those shown in Figure \ref{orbits}a. Close to the magnetopause loss cone, however, a large region is evident where particles exhibit significant transport characteristic of the orbits shown in Figure \ref{orbits}b and \ref{orbits}c. 
 
 For pitch angles at or close to 90 $^\circ$, the electrons exhibit large increases to both $\langle J \rangle$ and $\langle L \rangle$ whereas particles outside this region typically exhibit decreases to $\langle J \rangle$ and $\langle L \rangle$.
 This trend appears symmetric about 90 $^\circ$, as anticipated for the symmetric simulation conditions and possesses the same shape as the magnetopause loss cone boundary. The standard deviation across the bounce phase is in some instances comparable to the transport itself which highlights further complexities in trying to represent DOB transport in invariant space. In Figures \ref{Smap} and \ref{Nmap}, $\sigma$ is largest for larger pitch angles, most notably in Figure \ref{Smap} along two bands which run parallel to the loss cone. These represent regions where $\langle J \rangle$ and $\langle L \rangle$ can either increase or decrease, depending on the point in the electrons bounce phase that the drift shell bifurcates. There is also a notable minimum in the transport between 1.6 and 8 R$_E$ near 90 $^\circ$ for southward IMF.
 
 Several differences are apparent between Figures \ref{Smap} and \ref{Nmap}. The magnetopause loss cone is at higher L shells for the Northward IMF conditions due to the different shape and size of the dayside magnetopause. In both Figures, $\langle \Delta J \rangle$ and $\langle \Delta L \rangle$ peak inside the magnetopause boundary but for southward IMF, this effect is more pronounced and the red regions denoting increases to J and L, reflect the shape of the magnetopause loss cone. %
 Regions of large transport are also present directly adjacent to the loss cone which show the effects of current sheet scattering which can also cause particles to enter the atmospheric loss cone.

\subsection{Observable Signature}
 
 It is worth noting that the effects of DOBs are not easy to constrain via observations. The simulations predict a local depletion near the subsolar point of particles with small pitch angles and enhanced fluxes at higher latitudes, the latter of which indeed has been identified in several studies \citep{Antonova73,Antonova79,Antonova96,Savin98,Pissarenko01,Chen97,Chen98,Sheldon98}. Flux in the outer magnetosphere can, however, vary by orders of magnitude and this inherent variability makes it difficult to identify DOB transport predicted via theory and numerical simulations. 
 
\begin{figure*}[ht]
\centering
\includegraphics[width=0.65\textwidth]{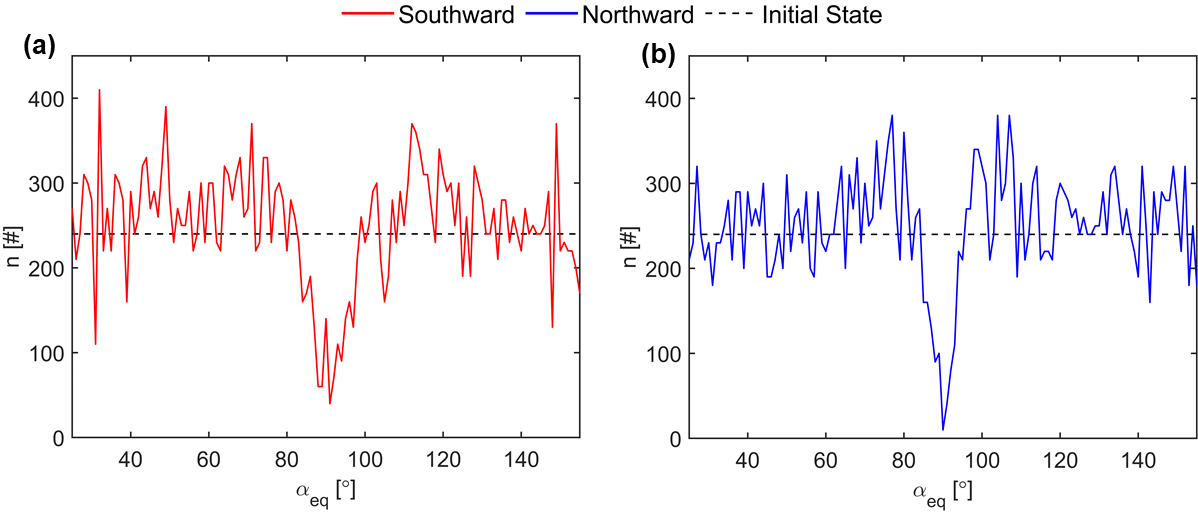}
\caption{Butterfly pitch angle distribution after 1 drift orbit for (a) southward and (b) northward IMF conditions .
\label{BPAD}}
\end{figure*}

 Figure \ref{BPAD} shows the equatorial pitch angle distribution on the night side corresponding to the maps shown in Figures \ref{Smap} and \ref{Nmap}. This shows that the consequence of large jumps in J for particles with small initial values of J, is net transport away from $\alpha_0$ = 90 degrees. 
  Butterfly PADs are observed throughout the night-time sector \citep{West73} and have been identified as deriving from drift shell splitting combined with magnetopause shadowing \citep[e.g.][]{Klida13}, interactions with ULF waves \citep[e.g.][]{Kamiya18} and particle injections \citep[e.g.][]{Artemyev15}. Here we identify DOBs as a mechanism which might also produce such a distribution. It is important to note, however, that the test-particle simulations presented herein do not contain kinetic effects such as the self-consistent generation and interaction with plasma instabilities. \citet{McCollough12} suggest that anisotropies can arise purely due to DOBs and these, as well as further plasma waves observed within the magnetosphere, might act to scatter particles and the dropouts near 90$^\circ$ therefore might be less pronounced than appear in Figure \ref{BPAD}.

\subsection{Radial Evolution}

In this section we examine the long term transport associated with the invariant changes displayed in Figures \ref{Smap} and \ref{Nmap}. The transport rates of relativistic electrons in the radiation belts are typically approximated by the Fokker-Plank equation where the second moment of the distribution function scales with time as $\langle (\Delta L)^2 \rangle = 2D_{LL}t$, where $D_{LL}$ is a diffusion coefficient which is applied in a drift averaged sense. This description is instantaneously utilised at multiple locations within the magnetosphere, such that a radial diffusion coefficient represents a Dirac delta function of particles at a particular radial distance, summed over azimuth. Within our simulations we see significant loss to the magnetopause over time which renders the resultant transport characteristics of an ensemble unphysical, and further complicates the scenario where the azimuthal third invariant is already undefined. \citet{Ukhorskiy14} examine the evolution of an ensemble with near 90 $^\circ$ pitch angles at a specific radial distance to examine the diffusion characteristics. Here, to study the effect of DOBs on outer belt electrons, we  choose to analyse the bulk transport characteristics of a distribution function which includes multiple L shells, to examine and isolate the net effect of DOBs on the outer radiation belt near the magnetopause.

\begin{figure*}[ht]
\centering
\includegraphics[width=1.0\textwidth]{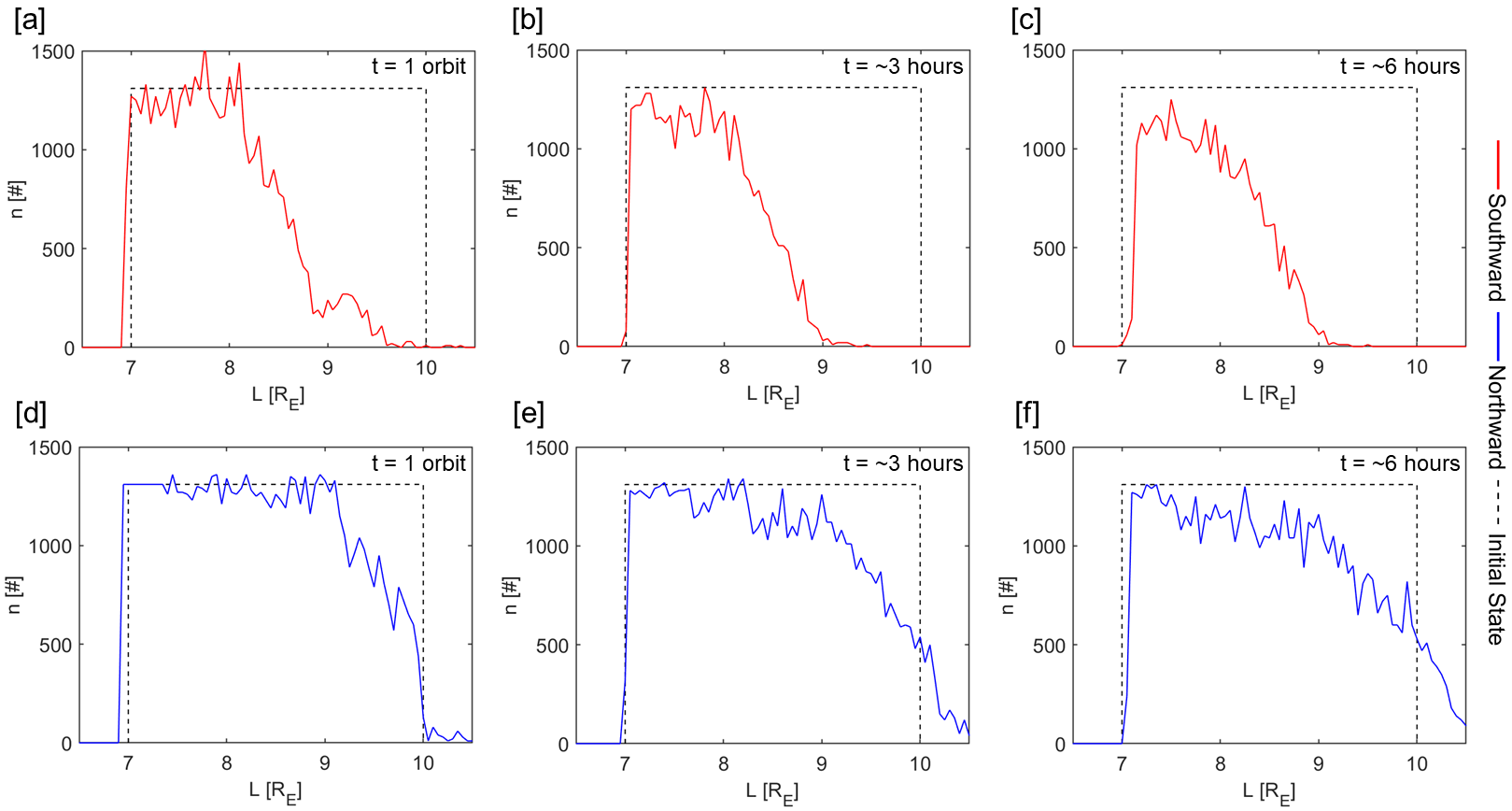}
\caption{Summed distribution function at three instances over 6 hours. Upper panels (a--c) shows the evolution under southward IMF conditions whereas lower panels (d--f) shows the evolution under northward IMF conditions.
\label{Lev}}
\end{figure*}

\begin{figure*}[ht]
\hspace{-4em}
\includegraphics[width=1.2\textwidth]{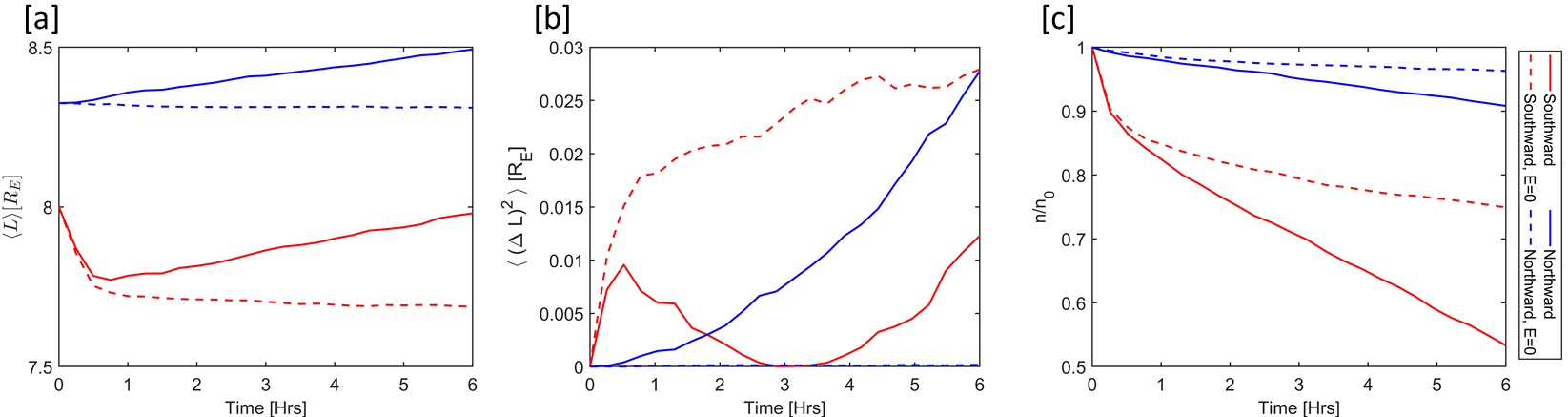}
\caption{(a) shows the mean location of the distribution function over 6 hours, (b) shows the corresponding mean change squared, $\langle (\Delta L) \rangle^2$ and (c) shows number of particles remaining in the simulations normalised to the initial number following the initial drift orbit. 
\label{diffusion}}
\end{figure*}

Figure \ref{Lev} shows the radial evolution of the ensemble which encompasses all pitch angles and bounce phases. The three different times correspond to after 1 drift orbit, after 3 and then 6 hours where the particles have eventually undergone 25 or more drift orbits depending on their L shell. The initially flat distribution functions exhibits an initial decrease at higher L shells following a single drift orbit which corresponds to the loss cones shown in  Figure \ref{Smap} and \ref{Nmap}. The distribution is also no longer flat and features several peaks and troughs. Following this, further losses occur over time, most notably for southward IMF conditions. %
Figure \ref{diffusion} shows further characteristics of the radial distribution function over time, starting with the distributions shown in Figures \ref{Smap} and \ref{Nmap}. Also shown are the same particle simulations run with the electric field set to zero in order to constrain the transport characteristics associated purely with the non-dipolar magnetic structure of the global fields. This therefore removes the effect of the electric fields and allows comparison with prior results obtained from empirical and semi-empirical magnetospheric fields \citep{Ozturk07,Wan10,Ukhorskiy11,Ukhorskiy14}. 

We first concentrate on the scenarios where we include the full magnetospheric magnetic and electric fields, shown by the solid lines. In Figure \ref{diffusion}a, for southward IMF, the mean L shell of the entire distribution, $\langle L \rangle $, initially moves inward due to further initial losses to the magnetopause on the drift orbits following the initial one and then slowly moves outward over the next 6 hours. For northward IMF, the distribution does not encounter the same immediate losses but also moves steadily outward. This trend is found during multiple MHD timesteps and is thus attributed to the convective electric field as opposed to a $\delta B$ component induced by ultra-low-frequency fluctuations.
Over this time period Figure \ref{diffusion}b shows the squared radial change, $\langle  (\Delta L)^2 \rangle $. For the southward IMF conditions, the magnetopause losses dominate over the first half of the simulations and produce decreases of this parameter up to 3 hours after which this then exponentially increases. For northward IMF conditions this exponential increase is immediately apparent. Figure \ref{diffusion}c shows that during this time period losses to the magnetopause are significant but more so for the southward IMF conditions where nearly half of the particles are lost after 6 hours.

These scenarios display notable differences when the electric field is excluded, as shown by the dashed lines. In these scenarios, Figure \ref{diffusion}a shows the average of the distributions maintain a near-constant value, although for southward IMF there is a small decline in this parameter. For northward IMF, $ \langle \Delta L \rangle$ is less than 10$^{-4}$ but for the southward case this initially rises sharply which is followed by an approximately linear increase over the next 5.5 hours. This is mirrored in the magnetopause losses which appear minimal for northward IMF but are more pronounced for southward IMF where over 20\% of the particles are eventually lost. The northward IMF scenario, where the radial distribution stays approximately constant and does not diffuse over time, is in qualitative agreement with previous studies \citep[e.g.][]{Ukhorskiy11}, where particles can undergo DOBs for extended time periods and traverse back and forth in the magnetosphere as their second invariant and radial locations increase and decrease. There is a small net outward trend, however, and for southward IMF the simulations clearly demonstrate that the net effect of DOBs is outward radial transport and continual loss to the magnetopause.

When the particles see both B and E, the evolution, as shown, appears super-diffusive, i.e. $(\Delta L)^2 \propto t^a$ where a$>1$, and the discussion of the evolution in terms of diffusive behaviour is also problematic due to the finite and varying number of particles in the simulation and the aforementioned large influence of the convective electric fields. 
There is currently significant interest in understanding the limits of the diffusive regime \citep[e.g.][]{Omura09,Allanson20,Trotta20} where the advection of ensembles of particles, not included in traditional derivations of the Fokker-Planck equation, take hold. 
The self-consistent  global MHD and test particles simulations appear to indicate that convective and advective effects are significant for DOBs but also that there appears to be some underlying diffusive behaviour associated with particles encountering this non-dipolarity in the outer magnetosphere, as shown when the E field is set to zero for southward IMF and the consistent magnetopause losses.

\section{Discussion and Conclusions \label{conclusions}}

In this study we have developed test particle simulation integrated within Gorgon global MHD simulations and demonstrated that we can capture DOB dynamics of the outer radiation belt. Through a combination of particle trajectory and magnetic field line tracing, we were able to measure particle transport to a high level of precision and used this to constrain radial transport resulting from DOBs. The overall transport characteristics broadly agree with previous studies using semi-empirical Tsyganenko fields that this phenomenon affects particles within an Earth radius of the magnetopause loss cone and can induce outward or inward transport. Several complexities were, however, revealed within this region due to the dynamic nature of the MHD simulations and inclusion of self-consistent electric fields.

 The MHD simulations reproduced an equatorial field maximum comparable to semi-empirical Tsyganenko magnetic field models and preliminary convergence tests determined that a grid resolution of 1/4 R$_E$ was sufficient to resolve this magnetic structure. Through the analysis of individual trajectories, the increase or decrease to an electron's pitch angle when entering and exiting the resultant high latitude pocket of magnetic field minimum, was shown to directly  correlate with an increase or decrease in radial distance. This transport was then further constrained for ensembles of electrons for both southward and northward IMF orientations and the resultant transport was mapped as a function of pitch angle and L shell. The instantaneous transport rates displayed notable variations within this region with the most significant transport observed for electrons with small initial values of the second invariant occurring away from the magnetopause surface. This was surprising as the equatorial maximum in the magnetic field appeared strongest closest to the magnetopause boundary. 
The transport maps obtained for southward IMF conditions also displayed notable structure with two distinct bands of increased transport occurring inward from the magnetopause surface and running parallel to the pitch angle dependent magnetopause loss cone. These differences represent complexities to DOBs which aren't apparent from initial considerations of the global field topology. 

The long term transport resulting from DOBs appeared to be strongly dependent on the convective electric fields contained within the global fields, for both southward and northward IMF conditions. Including the electric fields appeared to induce net outward transport of particles orbiting within an Earth radius of the magnetopause loss cone. Several phenomena are capable of inducing dynamics effects close to the magnetopause boundary, even during steady solar wind conditions, such as local plasma motions near the magnetopause resulting from reconnection \citep{Eastwood15}, or Kelvin-Helmholtz surface waves \citep{Guo10}. The resultant transport consequently did not follow the diffusion approximation where $(\Delta$L)$^2$ $\propto$ t. To further understand this we excluded the electric field to examine the effects of the particles interacting purely with the magnetic fields. These results also displayed differences between southward and northward IMF conditions and the southward IMF conditions notably induced transport which did scale as $(\Delta$L$)^2$ $\propto$ t and indicates an underlying diffusive aspect to this phenomenon.

Predicting radiation belt dynamics using diffusive models typically means competing effects have to be treated individually \citep{Glauert05,Subbotin09}. The simulations reported herein confirms the difficulties in incorporating DOBs into this approach, namely due to the non-diffusive nature of the transport and the large variations in the transport as a function of the electrons bounce phase. The long-term transport did, however, display significant losses to the magnetopause which remained reasonably constant for all scenarios examined. This net loss to the magnetopause and atmospheric loss cones highlights that the idea of a last-closed-drift-shell is more complex than an instantaneous location coinciding with the magnetopause loss cone \citep{Albert18} but also presents a potential method for incorporating the net effects of DOBs into Fokker-Planck-type models via a loss term acting near the magnetopause.  

It was noted that despite observations of trapped populations in the high-latitude cusps \citep{Antonova73,Antonova79,Antonova96,Savin98,Pissarenko01,Chen97,Chen98,Sheldon98}, the effects of DOBs are not easy to constrain via observations and it remains unclear to what extent DOBs persist and influence the highly variable outer radiation belt. We therefore identify butterfly pitch angle distributions in the night-time sector as a further observational signature which may be observed and linked to DOBs.

\section*{Acknowledgements}
RTD, JPE and JPC acknowledge funding from NERC grant NE/P017347/1 (Rad-Sat). HJA is supported by an Alexander Von Humboldt fellowship. OA is supported by a NERC Independent Research Fellowship NE/V013963/1. CEJW acknowledges grants NE/V002759/2 and ST/W002078/1. MOA holds a UKRI (STFC / EPSRC) Stephen Hawking Fellowship EP/T01735X/1. JWBE is supported by Grant NE/P017142/1 (SWIGS). RH, SG and NM would like to acknowledge the Natural Environment Research Council Highlight Topic grant NE/P10738X/1 (Rad-Sat) and the NERC
grants NE/V00249X/1 298 (Sat-Risk) and NE/R016038/1. This work
used the Imperial College High Performance Computing Service (doi: 10.14469/hpc/2232).

\bibliographystyle{abbrvnat}
\bibliography{main}

\end{document}